\documentclass{caosp302}
\usepackage{float}
\usepackage{graphicx}

\articleNo{123}
\pubyear{2007}
\volume{35}
\volnumber{3}
\firstpage{1}
\received{October, 2007}
\accepted{August 28, 2005}

\begin{document}

%
\htitle{Magnetic fields in Herbig~Ae stars}
\hauthor{R.V.\,Yudin, M.A.\,Pogodin, S.\,Hubrig and M.\,Sch\"oller}

\title{Circumstellar magnetic fields in Herbig~Ae stars}


%
\author{
       R.V. Yudin\inst{1}
       \and  
       M.A. Pogodin\inst{1}
       \and  
       S. Hubrig\inst{2}
      \and 
        M. Sch\"oller\inst{2}
       }

%
\institute{
           Central Astronomical Observatory of the Russian Academy of Sciences at Pulkovo,
           196140 Saint-Petersburg, Russia \email{ruslan61@hotmail.com}
            \and 
             ESO, Casilla 19001, Santiago 19, Chile 
         }

\date{October 8, 2007}

\maketitle

\begin{abstract}
We present the results of our latest studies of the circumstellar magnetic fields in Herbig~Ae stars 
and briefly discuss the cause of the failure of another recent study by our colleagues to 
confirm the Zeeman features in our spectra.
\end{abstract}

%
\section{Observations and results}
Numerous theoretical works predict the presence of a global
magnetic field of a complex configuration around Herbig~Ae stars.
Our longitudinal magnetic field determinations for these stars were 
obtained with FORS\,1 at the VLT in service mode from April 2003 to June 2005 at a resolution of 
2000 to 4000.
A magnetic field at a level higher than 3$\sigma$ was diagnosed for the stars 
HD\,31648, HD\,139614, and HD\,144432 (Hubrig et al.\ 2004, 2006, 2007).
We also showed that the Ca\,\textsc{ii} H\&K lines in the Stokes~V/I
spectra of the Herbig~Ae stars HD\,31648 and HD\,190073 display multi-component complex structures. 
These lines are very likely formed at the base of the stellar wind, as well as in 
the accretion gaseous flow. In our studies we concluded that a magnetic field is present in both stars, 
but it is mostly of circumstellar origin.
Using only the Ca\,\textsc{ii} H\&K lines for the measurement of circular polarization in HD\,190073, 
we were able to diagnose a longitudinal magnetic field at 2.8$\sigma$ level, $\left<B_z\right>=+84\pm30$ G.
This value is in full agreement with the high resolution spectropolarimetric data obtained 
with ESPaDOnS (Catala et al.\ 2007) who measured a longitudinal magnetic field
$\left<B_z\right>=+74\pm10$\,G using metallic lines. Unfortunately, the S/N ratio of the ESPaDOnS 
polarimetric spectra is too low in the spectral region containing the Ca\,\textsc{ii} doublet, and thus 
it was not used to measure the circumstellar magnetic field.

\begin{figure}
\begin{center}
\includegraphics[width=0.45\textwidth,angle=0,clip=]{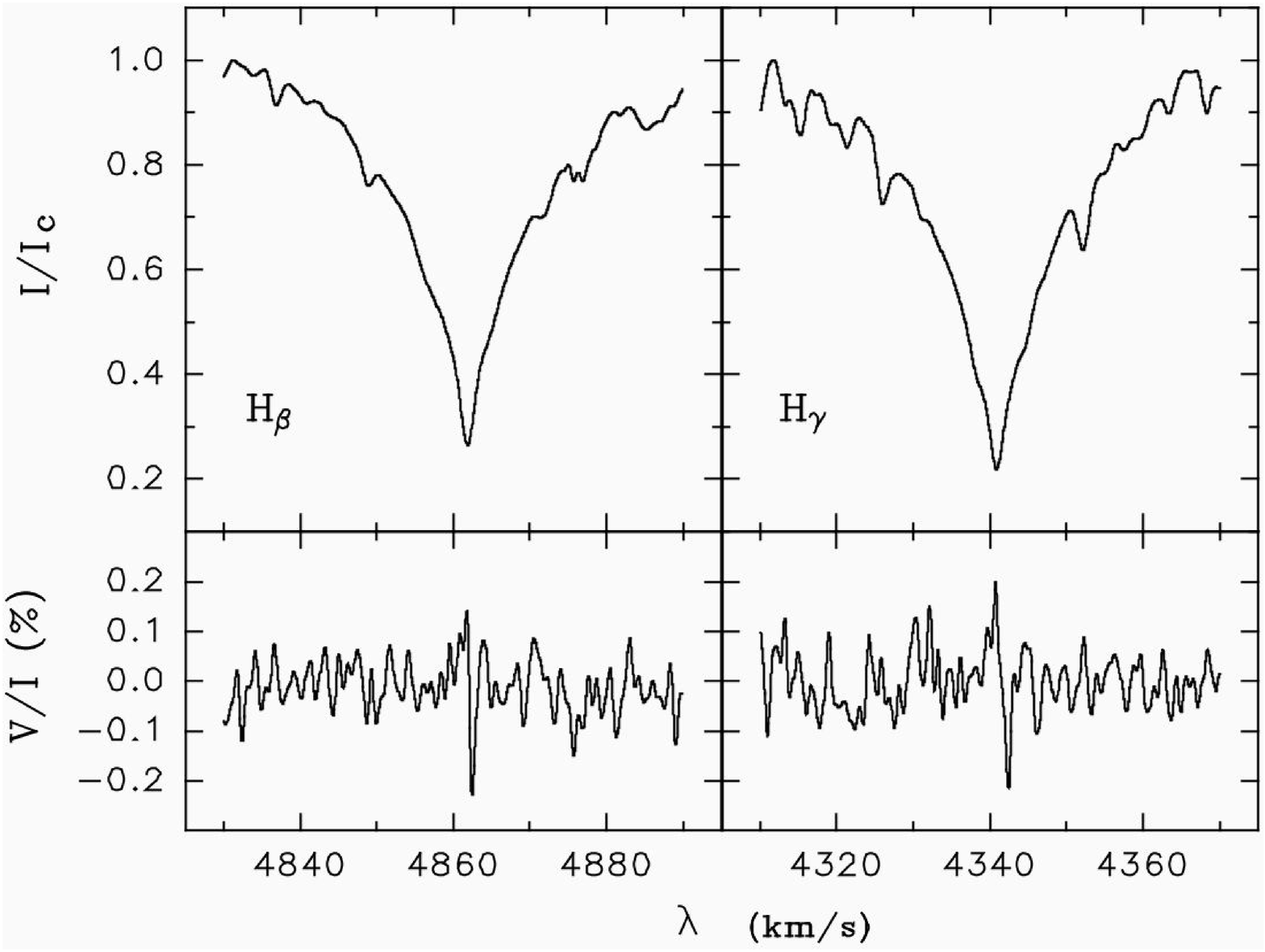}
\includegraphics[width=0.45\textwidth,angle=0,clip=]{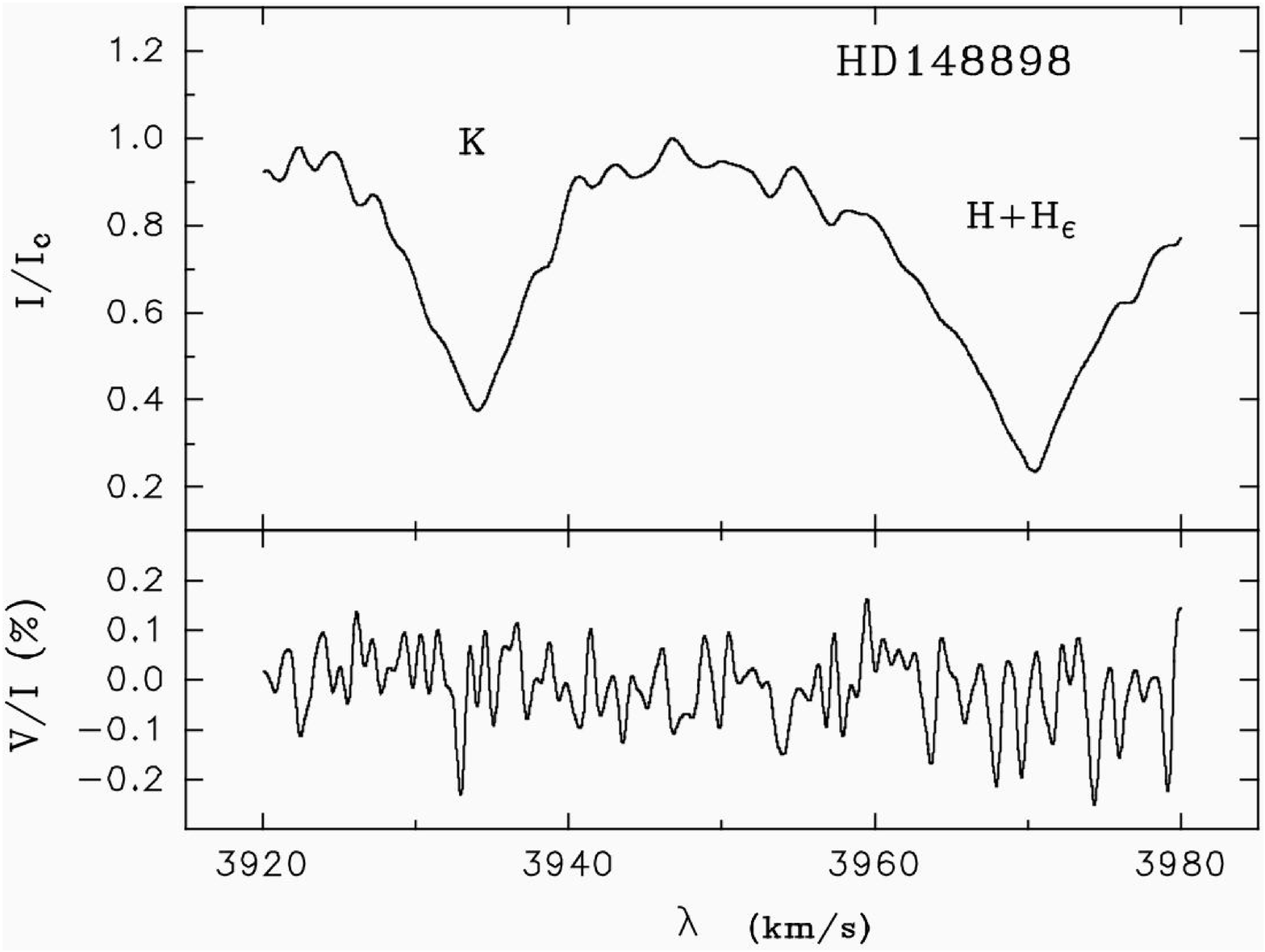}
\end{center}
\caption{
The Ca\,\textsc{ii} H\&K profiles (left panel)
   and H$\beta$ and H$\gamma$ profiles (right panel) of HD\,148898 and the corresponding
   Stokes~V spectra.
}
\label{fig:he3}
\end{figure}

\section{Discussion}
Recently, Wade et al. (2007) reported that their FORS\,1 spectra of the same Herbig~Ae stars
do not show Stokes~V signatures in the Ca\,\textsc{ii} H\&K lines.
They noted that their measurements obtained using broad slit widths, up to 1\arcsec{}, greatly degrade 
the resolution of the Stokes~V spectra causing the wrongly deduced longitudinal magnetic fields. 
Further, the authors claimed that they were not able to find polarization features 
in our FORS\,1 spectra, which they had downloaded from the ESO archive, either.
Quite puzzled by their report and 
even more by their apparent problem with the reduction of FORS\,1 spectra we compared  our measurements 
of Herbig~Ae stars with those of other A-type stars.
In Fig.~1 we present an  example of Stokes~I and Stokes~V spectra of the weakly magnetic
A-type star HD\,148898 with strong Ca\,\textsc{ii} H\&K lines observed
by us with FORS\,1 with a spectral resolution of 2000. Our
reduction has been carried out in the same way as for Herbig~Ae stars.
No Zeeman features could be detected at the positions of the Ca\,\textsc{ii} doublet,
confirming that the presence of distinct Zeeman features in Herbig~Ae stars is not a reduction artefact.
However, noticeable Zeeman features appear at the position of the hydrogen H$\beta$ and H$\gamma$ lines.
It is not clear to us why Wade et al.\ did not find the polarisation Ca\,\textsc{ii} H\&K features we have repeatedly found in Herbig~Ae
stars, but not in other stars with strong Ca\,\textsc{ii} H\&K lines.
Since we do not know their exact data reduction software, we can neither blame nor rule out a problem with their routines.

{}

\begin{thebibliography}{}


\article{Catala, C., Alecian, E., Donati, J.-F., Wade, G. A., Landstreet, J. D., B\"ohm, T., Bouret, J.-C., Bagnulo, S., Folsom, C., Silvester, J.}
{2007}{\aaa}{462}{293}

\article{Hubrig, S., Sch\"oller, M., Yudin, R.V.}{2004}{\aaa}{428}{L1}

\article{Hubrig, S., Yudin, R.V., Sch\"oller, M., Pogodin, M.A.}{2006}{\aaa}{446}{1089}

\article{Hubrig, S., Pogodin, M.A., Yudin, R.V., Sch\"oller, M., Schnerr, R.S.}{2007}{\aaa}{463}{1039}

\article{Wade, G.A., Bagnulo, S., Drouin, D., Landstreet, J.D., Monin, D.}{2007}{\mnras}{376}{1145}

%
%
%
%













\end{thebibliography}
\end{document}